\documentclass{article}

\usepackage{PRIMEarxiv}

\usepackage[utf8]{inputenc} 
\usepackage[T1]{fontenc}    
\usepackage{hyperref}       
\usepackage{url}            
\usepackage{booktabs}       
\usepackage{amsfonts}       
\usepackage{nicefrac}       
\usepackage{microtype}      
\usepackage{lipsum}
\usepackage{fancyhdr}       
\usepackage{graphicx}       
\graphicspath{{media/}}     
\usepackage{caption}
\usepackage{subcaption}
\title{Study of Kinematics of the Jets in VBS Process at the Large Hadron Collider
}

\author{
  Kadri Ozdemir\\
  Department of Fundamental Sciences \\
  İzmir Bakırçay University \\
  Menemen, İzmir\\
  \texttt{ kadri.ozdemir@bakircay.edu.tr}\\
}

\begin{document}
\maketitle

\begin{abstract}
The comparison of the kinematic properties of different Vector Boson Scattering (VBS) and Vector Boson Fusion (VBF) processes using multivariate discriminator is presented. The search is performed to identify common features in the the polarized $W^{\pm}W^{\pm}jj$ channel, mainly central and forward region jets kinematic variables such as eta, phi, mass and transverse energy, in order to determine some universal cuts. Traditionally boosted decision trees (BDT) are used as  multivariate discriminator. The presented results are based on proton-proton collision Monte Carlo (MC) simulation samples  center-of-mass energy 14 TeV and corresponding to an integrated luminosity of 3000 $fb^{-1}$. MC samples are generated using MADGRAPH and PYTHIA. 
\end{abstract}


\section{Introduction}
The inmost structure of the Standard Model (SM) electroweak (EW) interactions is examined by VBS. Vector Boson Scattering supply solitary sensitivity to new physics phenomena in the gauge sector. In this reason, VBS and VBF  are important processes to understand SM EW sector that is well established in high energy physics\cite{1}.The longitudinal modes of the weak bosons are appearances of the Nambu-Goldstone bosons, which arise from the spontaneous breaking of the EW symmetry, at the very high  scattering energies. In order to  help reveal the dynamics  behind the Higgs mechanism, it is important to investigate VBS interaction. Furthermore, VBS is convient to testing the gauge structure of electroweak interaction because of their interplay with trilinear couplings resulting potentially large gauge cancellations and the contribution of quartic interactions. VBS process is unique with this properties. Therefore, VBS is crucial not only researching beyond-the Standard Model (BSM) scenarios comprehensively but also accomplishing precise predictions in the SM of high energy physics. The precise prediction in the SM and BSM scenarios are both related to specific models and effective field theory (EFT) frameworks\cite{2,3}.The importance  VBS physics can be found in \cite{4,5,6,7}.\par
\begin{figure}
	\centering
        \includegraphics[width=16cm]{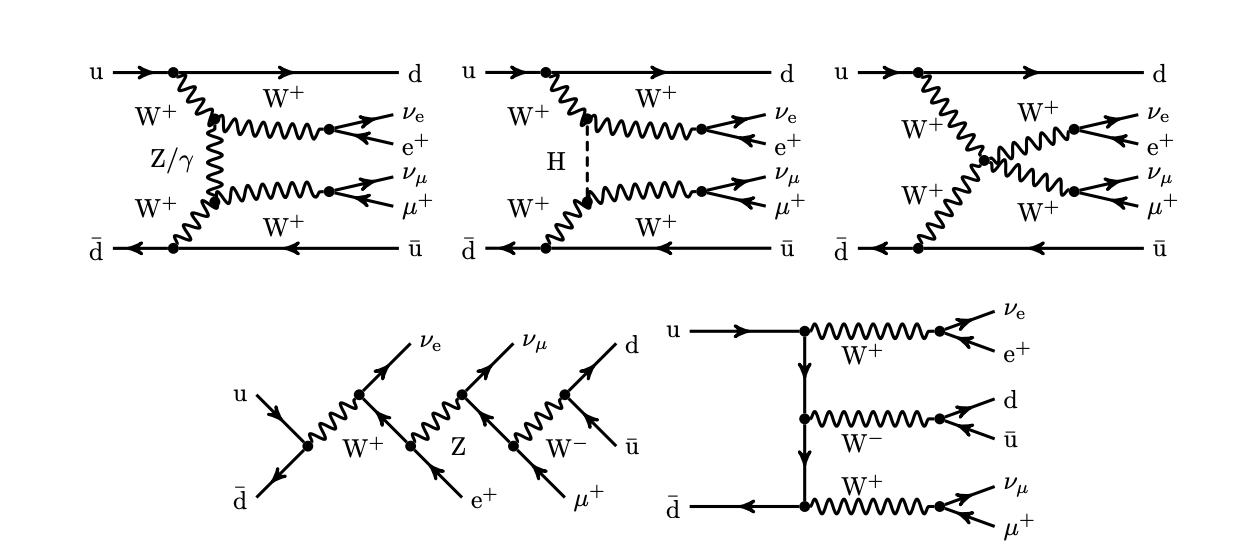}
	\caption{Feynman diagrams for VBS contributions (top) as well as non-VBS contributions (bottom). \cite{8}}
	\label{fig:fig1}
\end{figure}

VBS and VBF are important processes to study at the Large Hadron Collider (LHC) at CERN for both the ATLAS and the CMS experiments. A scattering process at the LHC is defined through its measured final state which consists of jets, leptons, photons or missing energy.The measurement of VBS at the LHC, a quark or anti-quark scatters with another quark or antiquark via a space-like exchange of two gauge bosons, called as  a W or Z boson or a photon \cite{1,3} (See Fig\ref{fig:fig1}).VBS process decay the product of heavy gauge bosons which has quarks in the initial state and at least two quarks and up to four leptons in the final state. VBS  has three possible signatures at the LHC. First signature is 4 leptons and 2 jets which is called fully leptonic decay. Second one is called as semi hadronic decay, including 2 leptons and 4 jets. Third one is 6 jets at final states which entitle as fully hadronic decay. The advantages of this definition are describing off shell, all non-resonant and interference effects and also being clearly gauge invariant. \cite{7,8}. The fully leptonic decay of VBS process is studied in this research which means two high energetic jets, produced in association with one or two vector bosons, respectively.\par
The polarized $W^{\pm}W^{\pm}jj$ production is studied at ATLAS and CMS experiments at the LHC which is decay with exactly two leptons, $l = e$ or ${\mu}$, missing transverse momentum induced by the neutrinos, and at least two jets with a large rapidity separation. This decay is a rare but also the key process to experimentally probe the Standard Model (SM) nature of electroweak symmetry breaking (EWSB) and the Higgs mechanism \cite{9,10}. $W^{\pm}W^{\pm}jj$ production is analyzed due to being an important benchmark process and  permitting the study of the jet activity in a relatively large event sample.\par
In this paper, section 2 is devoted to VBS processes for both ATLAS and CMS experiments at the LHC. The latest VBS results are discussed in this section. Section 3 is dedicated to the polarized $W^{\pm}W^{\pm}jj$ jets kinematic variables such as eta, phi, mass and transverse energy results and theirs universal cuts. The results of the polarized $W^{\pm}W^{\pm}jj$  channel jets kinematic variables and theirs universal cuts using BDT are presented section 3. 
\section{VBS at LHC}
\label{sec:headings}

The Large Hadron Collider (LHC) is a 27-kilometer ring of superconducting magnets with a series of accelerator structures to boost the energy of the particles along the way. The proton beams are collide inside the LHC at  around the accelerator ring, four locations  that  are placed ATLAS, CMS, ALICE and LHCb detectors. Whereas ATLAS and CMS  are desined for general purpose detectors such SM and BSM physics, LHCb and ALICE are designed for b-physics and heavy-ion physics respectively.\par
With the start of operations at the LHC in 2008, a new era in high-energy physics has begun. LHC was delivered 7 and 8 TeV center-of-mass energies protons between 2009-2013 to its experiments, which is referred as Run-I phase. The Run-I phase led not only to the discovery of the Higgs boson with a mass of 125 GeV \cite{11,12}, but also to the observation of the predicted particles from the SM and to determine the parameters of these SM particles. With the discovery of the Higgs boson, it became clear that the Standard Model is a complete and coherent theory, but it remain unexplained some phenomena such as matter-antimatter asymmetry, evidence of neutrino oscillations, dark matter and masses of elementary fermions. Therefore, extensions of the SM have been hypothesised. These extensions predict either parameter deviations from the SM predictions, or new particles, or both. Experimental searches and measurements to find signs of BSM probe many different scattering processes as in  ATLAS and CMS experiments at the LHC. VBS is certainly a pre-eminent one of the various processes that attract the attention of theorists and experimentalists. In fact, it searches both the gauge interactions and the couplings between the Higgs and the gauge bosons, that are two key aspects of the SM together. The gauge interactions are one of the few processes with tree-level sensitivity to the quadrilinear gauge couplings. The couplings between the Higgs and the gauge bosons are investigated at energy scales that can be meaningfully different from the Higgs mass \cite{13}. LHC Run-II period is between 2015-2018, proton beam colliding 13 TeV center-of-mass energies which was double the LHC Run-I energy. In LHC Run-II, the first evidence for rare VBS processes was collected by ATLAS and CMS experiments. Not only were advances made in experimental results, but several important advances were made in theory as well at the VBS. Therefore, one of the aims of this study is to present improvements that can be achieved with LHC Run-III by comparing the kinematic properties of different VBS process using multivariate discriminator for both ATLAS and CMS experiments.\par
Run-III data taking has just started for the experiments at the LHC. The new data taking period will continue for the LHC over the next four years, and the protons will be accelerated with an center-of-mass energies of $13.6$ TeV. Therefore, it will be important to probe rare VBS processes with the using new multivariate discriminator cuts. In order to better understand this study, the picture of the current state-of-the art in the VBS results obtained with the Run-II data from the CMS and ATLAS experiments should be summarized. Firstly, ATLAS VBS results will be given in this section. Afterwards, CMS VBS results will be discussed in detail.

\subsection{Current ATLAS results in VBS}
ATLAS detector is designed for multipurpose which is consist of inner detectors, magnets, calorimeters and muon spectroscopy. The inner detector, embedded in a 2 T solenoidal field, comprises semiconductor pixel, strip detectors and straw-tube tracking detectors; whereas in the inner part of the tracking both semiconductor pixel and strip detectors provides pattern recognition, momentum and vertex measurements, and electron identification, straw-tube tracking detectors  generate and detect transition radiation in its outer part. The calorimeter system, surrounded by muon system, of ATLAS detector consists of high granularity liquid-argon (LAr) electromagnetic sampling calorimeters and a scintillator-tile hadronic calorimetry, that both provide  electromagnetic and hadronic energy measurements \cite{14}.\par 
LHC Run-I phase, ATLAS experiment has examined VBS in the $W^{\pm}W^{\pm}$, $Z_{\gamma}$ and $W_{\gamma}$ final states in proton-proton collision data corresponding to 20.3 $fb^{-1}$ at $\sqrt{s}=$ 8 TeV \cite{15,16,17,18}. The electroweak production of $W^{\pm}W^{\pm}$ and  $WZ$ bosons, with the W and Z bosons decaying to leptons, and the electroweak production of $WZ$, $WW$ and $ZZ$ bosons, with one gauge boson decaying to leptons and the other decaying hadronically are studied at  $\sqrt{s}=$ 13 TeV  in the LHC Run-2. The dijet invariant mass distribution of the electroweak $W^{\pm}W^{\pm}jj$ production of the selected candidate events in the signal region is shown in Figure 2(a). Comparisons of theoretical predictions with various configurations of event generators and parton-shower programs is shown in Figure 2(b).
 During the LHC Run-II data-taking, ATLAS experiment has collected $\mathcal{L}$ = 140 $fb^{-1}$ at $\sqrt{s}=$ 13 TeV which leads to 
 not only first observation of $W^{\pm}W^{\pm}$, $WZ$ and $ZZ$ channel for leptonic decay modes but also the first study of VBS using semi-leptonic decays targeting WZ/WW events \cite{21,22}.

\begin{figure}[h]
\begin{subfigure}{0.5\textwidth}
\includegraphics[width=0.9\linewidth, height=8cm]{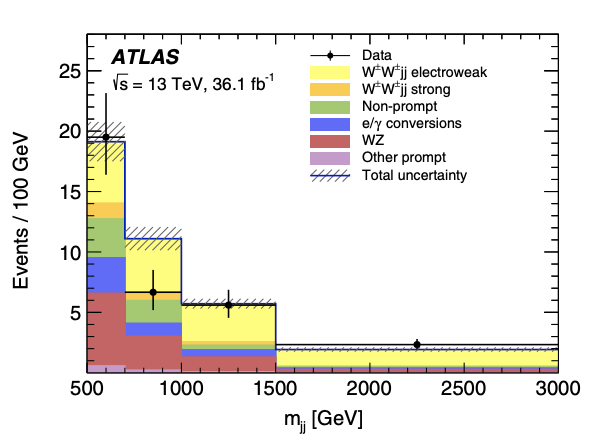}
\caption{}
\label{fig:subim1}
\end{subfigure}
\begin{subfigure}{0.5\textwidth}
\includegraphics[width=0.9\linewidth, height=8cm]{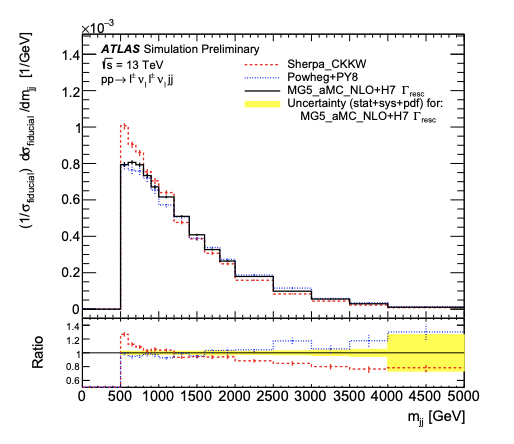}
\caption{}
\label{fig:subim2}
\end{subfigure}
\caption{(a) Dijet invariant mass distribution of the electroweak $W^{\pm}W^{\pm}jj$ production of the selected candidate events in the signal region \cite{19} (b)Comparisons of theoretical predictions with various configurations of event generators and parton-shower programs \cite{20}}
\label{fig:fig2}
\end{figure}

\subsubsection{Current CMS results in VBS}
The main feature of the CMS apparatus is a superconducting solenoid with an internal diameter of 6 m, which provides a magnetic field of 3.8 T. Inside the superconducting solenoid, a silicon pixel and strip tracker, a lead tungstate crystal electromagnetic calorimeter,  a brass and scintillator hadron calorimeter consists of a barrel and two end cap sections. The hadronic outer calorimeter which is designed to measure high energetic jets, and muon spectrometer are placed outside of the solenoid \cite{23}.\par
LHC Run-I phase, CMS Collaboration investigated firstly diboson production of the  VBS in the $W^{\pm}W^{\pm}$, $Z_{\gamma}$ and $W_{\gamma}$ final states in proton-proton collision data corresponding to 19.4 $fb^{-1}$ at $\sqrt{s}=$ 8 TeV \cite{24,25,26}. Later the electroweak production of $W^{\pm}W^{\pm}$ and  $WZ$ bosons, with the W and Z bosons decaying to leptons, and the electroweak production of $WZ$, $WW$ and $ZZ$ bosons, with one gauge boson decaying to leptons and the other decaying hadronically are studied at  $\sqrt{s}=$ 13 TeV  data corresponding to 35.9 $fb^{-1}$in the LHC Run-2. The dijet invariant mass distribution of the electroweak $W^{\pm}W^{\pm}jj$ production of the selected candidate events in the signal region is shown in Figure 3(a) The dilepton invariant mass distribution of the electroweak $W^{\pm}W^{\pm}jj$ production of the selected candidate events in the signal region is shown in Figure 3(b). The distribution of the BDT output in the control region obtained by selecting $W^{\pm}W^{\pm}jj$ events is given in Figure 4.\par
CMS apparatus collected $\mathcal{L}$ = 137 $fb^{-1}$ at $\sqrt{s}=$ 13 TeV at the LHC Run-III data-taking, the experimental outlook for searches and measurements of VBS processes is very bright as in ATLAS experiment. With these collected data, not only the $WZ$ and $W^{\pm}W^{\pm}$ processes have been measured and observed differentially but also the measurements of the the VBS ZZ production with leptonic decays was also updated which is exceeding the $3\sigma$ threshold for evidence but  not enough to reach $5\sigma$ observation \cite{27,28,29,30}.

\begin{figure}[ht]
\begin{subfigure}{0.5\textwidth}
\includegraphics[width=0.9\linewidth, height=8cm]{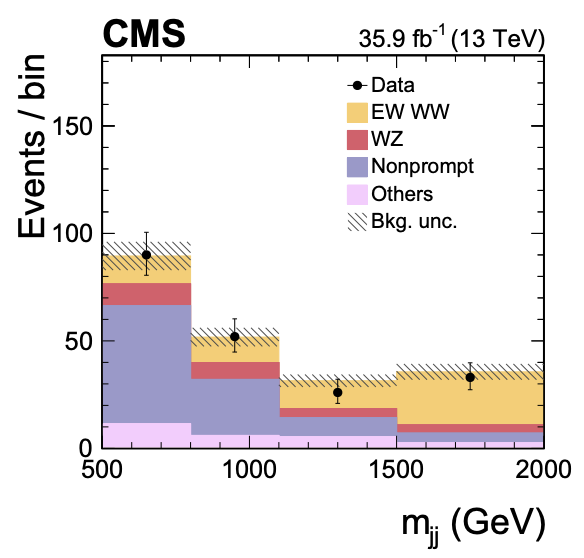}
\caption{}
\label{fig:subim3}
\end{subfigure}
\begin{subfigure}{0.5\textwidth}
\includegraphics[width=0.9\linewidth, height=8cm]{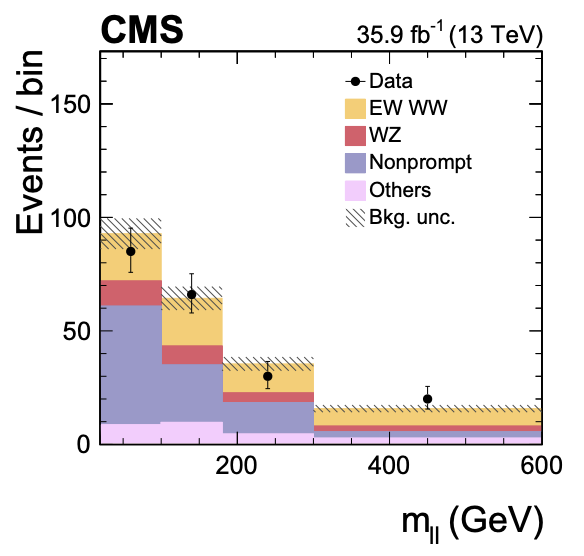}
\caption{}
\label{fig:subim4}
\end{subfigure}
\caption{(a) Dijet invariant mass distribution of the electroweak $W^{\pm}W^{\pm}jj$production of the selected candidate events in the signal region \cite{27} (b) Dilepton invariant mass distribution of the electroweak $W^{\pm}W^{\pm}jj$production of the selected candidate events in the signal region \cite{27}}
\label{fig:fig3}
\end{figure}
\begin{figure}[ht]
	\centering
        \includegraphics[width=8cm]{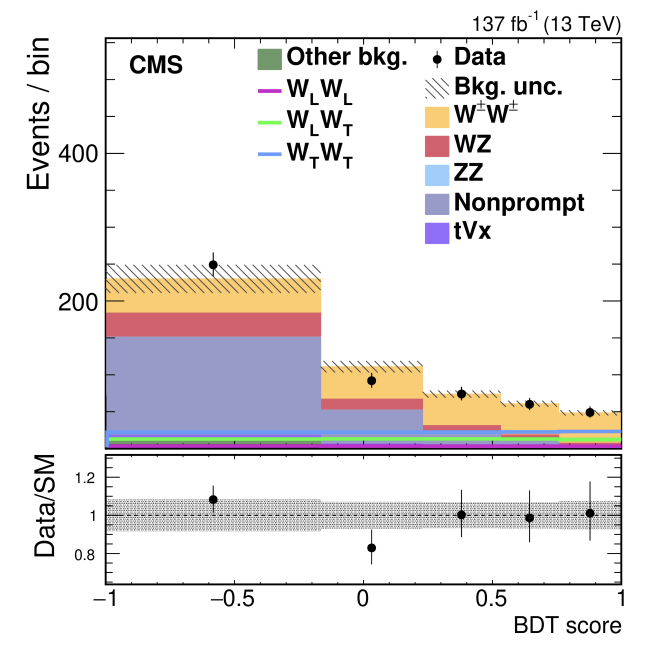}
	\caption{Distribution of the BDT output in the control region obtained by selecting $W^{\pm}W^{\pm}jj$ events \cite{27}}
	\label{fig:fig4}
\end{figure}

\section{Results and Discussion}
The same-sign  $W^{\pm}W^{\pm}jj{\rightarrow}l^{\pm}{\nu}l^{\pm}{\nu}jj$ process is considered to be the "golden" channel in the study of VBS because of low background and its distinctive signature. Compared to other diboson processes, it has the largest ratio of electroweak to strong production processes, since the strong production diagrams are strongly suppressed. In addition, purely electroweak diagrams, which are insensitive to the self interactions of the gauge-bosons, are compressed. The quantum chromodynamics (QCD)-induced production of $W^{\pm}W^{\pm}jj$ events is considerably sub-dominant according to the EW contribution in contrast to other VBS modes. It can also be seen from VBS production that the cross section ratio of EW components compared to QCD is very large, of the order of 4-6 in typical fiducial regions, whereas it is usually less than 1 for other processes. This is owing to charge conservation, which prohibits gluon-initiated processes in the QCD background, contrary to other VBS processes. Therefore, the $W^{\pm}W^{\pm}jj$ is the most convenient channel to search potential new-physics effects, including those affecting polarization and anomalous quartic gauge couplings. Both ATLAS and CMS experiments have measured and observed $W^{\pm}W^{\pm}jj$ channel at $\mathcal{L}$ = 140 $fb^{-1}$ at $\sqrt{s}=$ 13 TeV during LHC Run-II as mentioned in previous chapter. This measurement and observation were lead to perform  the first study of a VBS process separated by the polarization of the vector bosons at the LHC. The importance of the polarization of the $W^{\pm}W^{\pm}jj$ channel for both the CMS and ATLAS experiments at the LHC is increasing, and the study of the jet kinematics makes a significant contribution to understand of this channel main backgrounds. The search is performed to identify common features in the the polarized $W^{\pm}W^{\pm}jj$ channel, mainly central and forward region jets kinematic variables such as eta, phi, mass and transverse energy, in order to determine some universal cuts.\par
The polarized $W^{\pm}W^{\pm}jj$ events in MADGRAPH are generated, which is decay with exactly two leptons, $l = e$ or ${\mu}$, missing transverse momentum illustrated with the neutrinos, and at least two jets with a large rapidity separation for this analysis. Polarization means that its definition depends on the reference frame and is not a Lorentz invariant quantity. The polarization definition is the parton-parton center-of-mass frame and in the W boson pair center-of-mass frame. Hence the cross section of the polarized states measure independently using definition of polarization. A polarization definition based on projections on the LO decay-angle distributions . The polarization desciription on projections on the LO decay-angle distributions is reconciled to inclusive LO distributions  owing to the resonant vector-boson diagrams. It is valid only for a polarized vector boson, whereas it fails for the other situation such as large background, tight cuts or huge NLO corrections. The MadGraph5 aMC@NLO Monte Carlo Event Generator has automated the calculation of polarized cross sections for LO calculations. Automation procedure has completed by employing decay chains in the narrow-width approximation and including spin correlations via the Madspin package. For that reason the polarized $W^{\pm}W^{\pm}jj$ events are generated MadGraph5 aMC@NLO Monte Carlo Event Generator which version is 3.2.0. L(R) for left (right) helicity information is added the polarized $W^{\pm}W^{\pm}$ event generation. With this version, the polarized $W_{L}W_{T}$  events are generated, which is labelled as longitudinally polarized W boson.\par 
The longitudinally polarized vector bosons is important channel because of their coupling to BSM interactions at large scattering energies, that VBS can probe BSM. The disadvantage of  $W_{L}W_{T}$ is irreducible background. At this point, it is necessary to distinguish between SM QCD backgrounds and VBS events. $W_{L}W_{T}$ QCD background is generated  MadGraph5 aMC@NLO as well. Both $W_{L}W_{T}$ signal and backgrounds are generated at $\sqrt{s}$ = 14 TeV  pp collision and corresponding to an integrated luminosity of 3000 $fb^{-1}$ at Run-III scenario. PYTHIA 6.4 and PYTHIA 8.3 are used as MC generation for parton showering, hadronization and producing nonstable hadrons decay the final particles. ATLAS and CMS official parton distribution functions (PDFs) CTEQ6L1 and NNPDF30 are used  respectively in this analysis for both $W_{L}W_{T}$ signal and background. The composite Higgs models is implemented to the event generation as a BSM features. The BSM benchmark scenario $a = 0.8$ and the SM  $a = 1.0$ are chosen in the event generation, where a is identified with the model parameter in Madgrapgh. The selection cuts in the below are applied  to boost VBS topology: 
\begin{itemize}
\item $p_{T}(jets)> 10 GeV $ 
\item $|{\eta}(jets)|< 5.0 $ 
\item $p_{T}(W^{\pm})> 20 GeV$
\item $|{\eta}(W^{\pm})|< 2.5 $ 
\item $m(jj)> 200 GeV$ 
\item $|{\Delta\eta}(jj)|< 2.5 $ 
\item $m(jj)> 200 GeV$ 
\item $m(W^{\pm}W^{\pm})> 250 GeV$ 
\end{itemize}
The effective cross section for $W_{L}W_{T}$ configuration for the composite Higgs (CH) scenario $a = 0.8$ and the SM  $a = 1.0$ is shown. Figure 5(a) shows first leading jet $p_{T}$ for particle level which are reconstructed anti-$k_{T}$ algorithms with cone size $R= 0.8$. NNPDF30 shows better trends than CTEQ6L1 for both Pythia8 and Pythia6 CMS and ATLAS detectors for first leading jet. Second leading jet $p_{T}$ for particle level is illustrated with Figure 5(b) which is show same trend as first leading jet. First leading jet $\eta$ distribution indicate at Figure 6 that particle level jet more spread in forward region at the ATLAS and CMS detectors. The generated events are reconstructed at ATLAS and CMS detectors by using DELPHES. The first and second leading jet $p_{T}$ distribution can be found in Fig.7(a) and Fig.7(b) for detector level. There is a good agreement between particle and reconstructed level.\par 
The search is performed to identify common features in the the polarized $W_{L}W_{T}$ channel, mainly central and forward region jets kinematic variables such as eta, phi, mass and transverse energy, in order to determine some universal cuts. Traditionally boosted decision trees (BDT) are used as  multivariate discriminator. ROOT TMVA program is used for multivariate discriminator. Traditionally BDT are used as multivariate discriminator and its inputs are $m(jj)$, $|{\Delta\eta}(jj)$, $m(W^{\pm}W^{\pm})$, zeppenfeld of two W boson, ratio between the transverse momentum of the jet system and the scalar $p_{T}$ sum of jets. After applying BDT, $W_{L}W_{T}jj$ cross-section result is improved $0.5-1.0{\%}$ for both CMS and ATLAS experiments at  $\sqrt{s}$ = 14 TeV  pp collision 3000 $fb^{-1}$ at Run-III scenario. 

\begin{table}
	\caption{The effective cross section for $W_{L}W_{T}$ configuration for the composite Higgs (CH) and SM scenario}
	\centering
	\begin{tabular}{||c c c c c c||}
		\toprule
		Process    & SM ${\sigma} (fb^{-1})$  & SM $(f_{\Lambda,\Lambda^{-1}})$ & CH ${\sigma} (fb^{-1})$  & CH $(f_{\Lambda,\Lambda^{-1}})$  & SM/CH \\
		\midrule
		$W_{L}W_{T}$ & 250.1  & 14.4  & 260.4  & 14.8 & 1.02\\
		
		\bottomrule
	\end{tabular}
	\label{tab:table}
\end{table}

\begin{figure}[ht]
\begin{subfigure}{0.5\textwidth}
\includegraphics[width=0.9\linewidth, height=9cm]{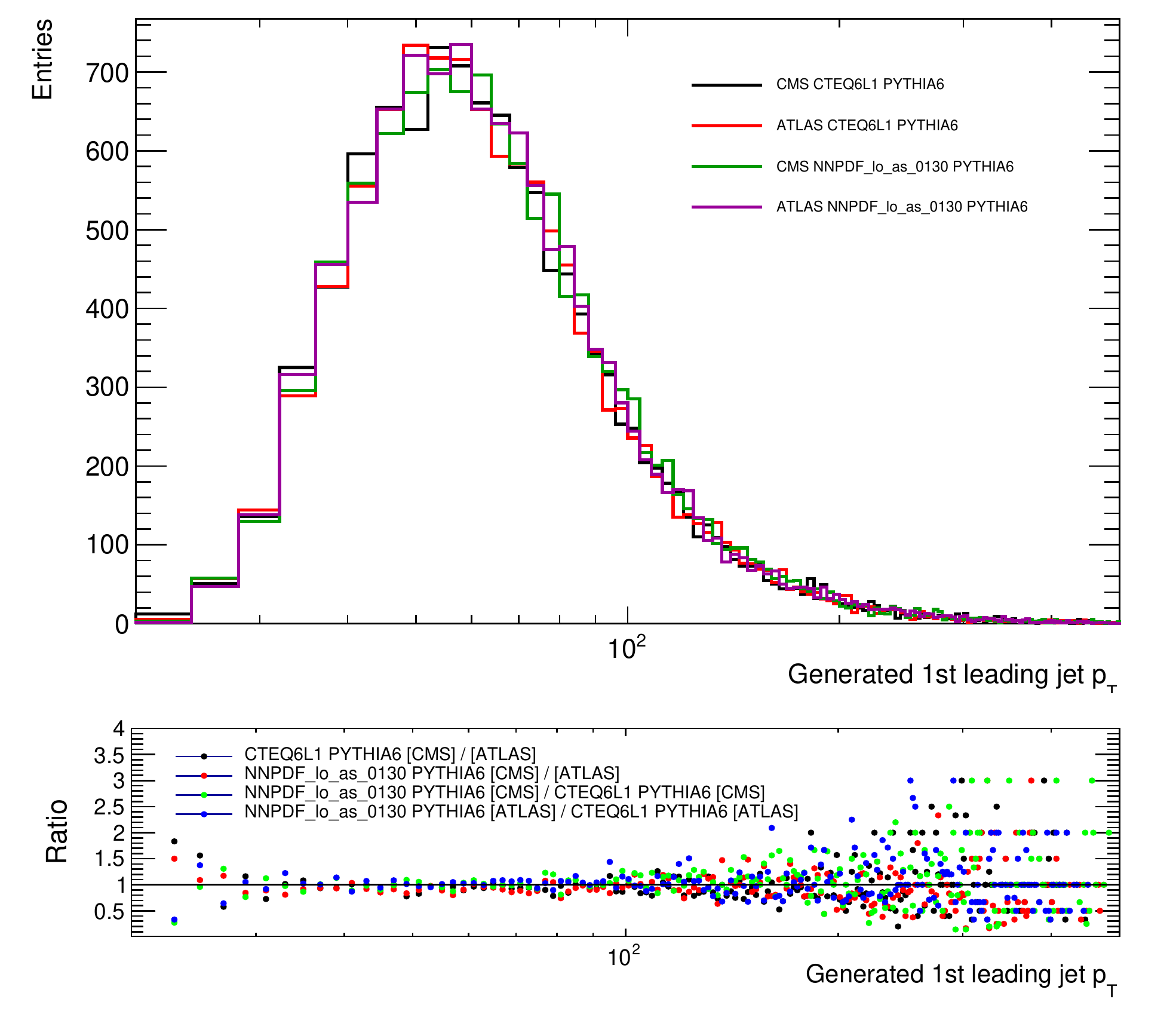}
\caption{}
\label{fig:subim5}
\end{subfigure}
\begin{subfigure}{0.5\textwidth}
\includegraphics[width=0.9\linewidth, height=9cm]{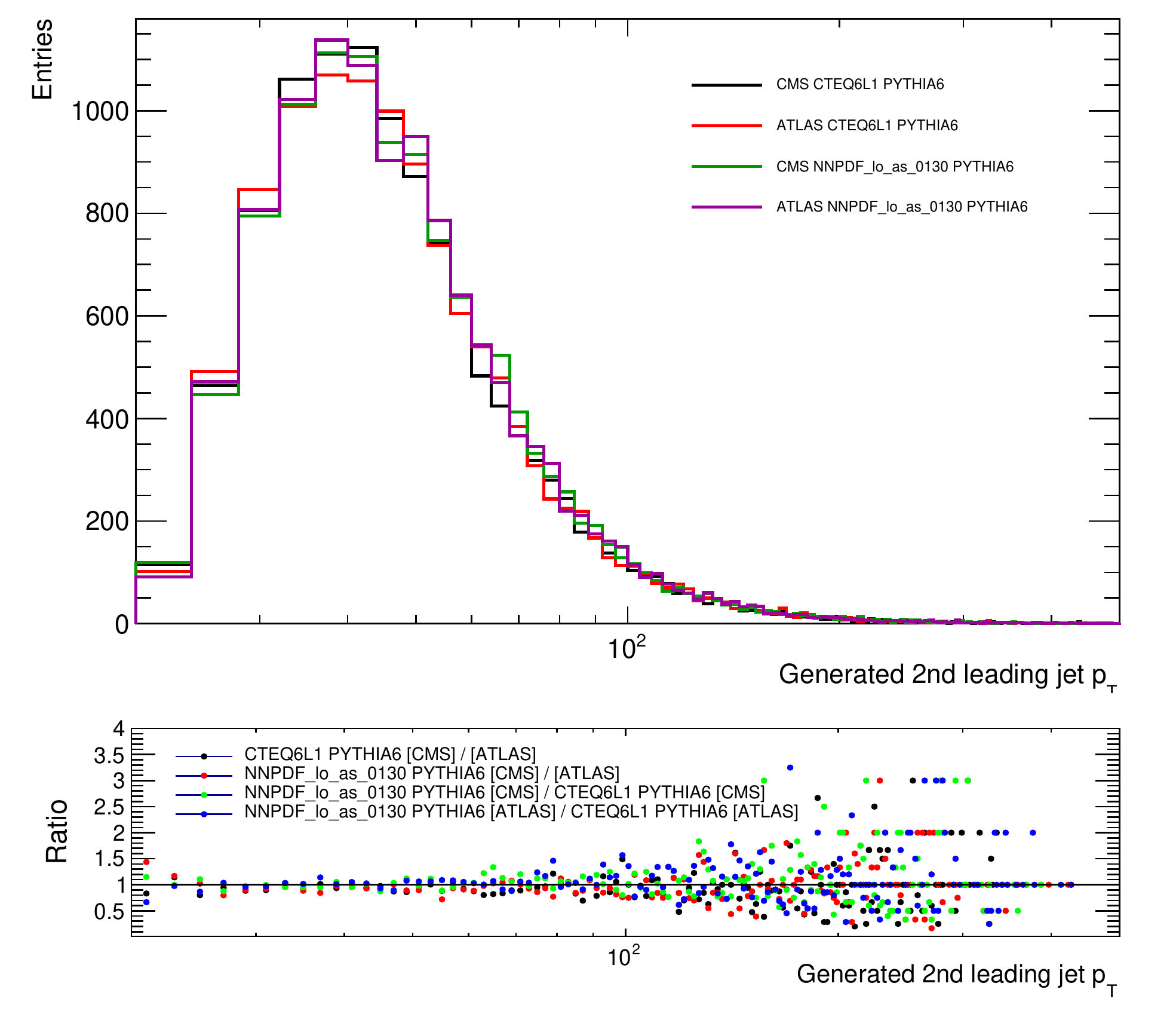}
\caption{}
\label{fig:subim6}
\end{subfigure}
\caption{(a)First leading jet $p_{T}$ for particle level which are reconstructed anti-$k_{T}$ algorithms with cone size $R= 0.8$ (b)Second leading jet $p_{T}$ for particle level}
\label{fig:fig5}
\end{figure}

\begin{figure}[ht]
	\centering
        \includegraphics[width=9cm]{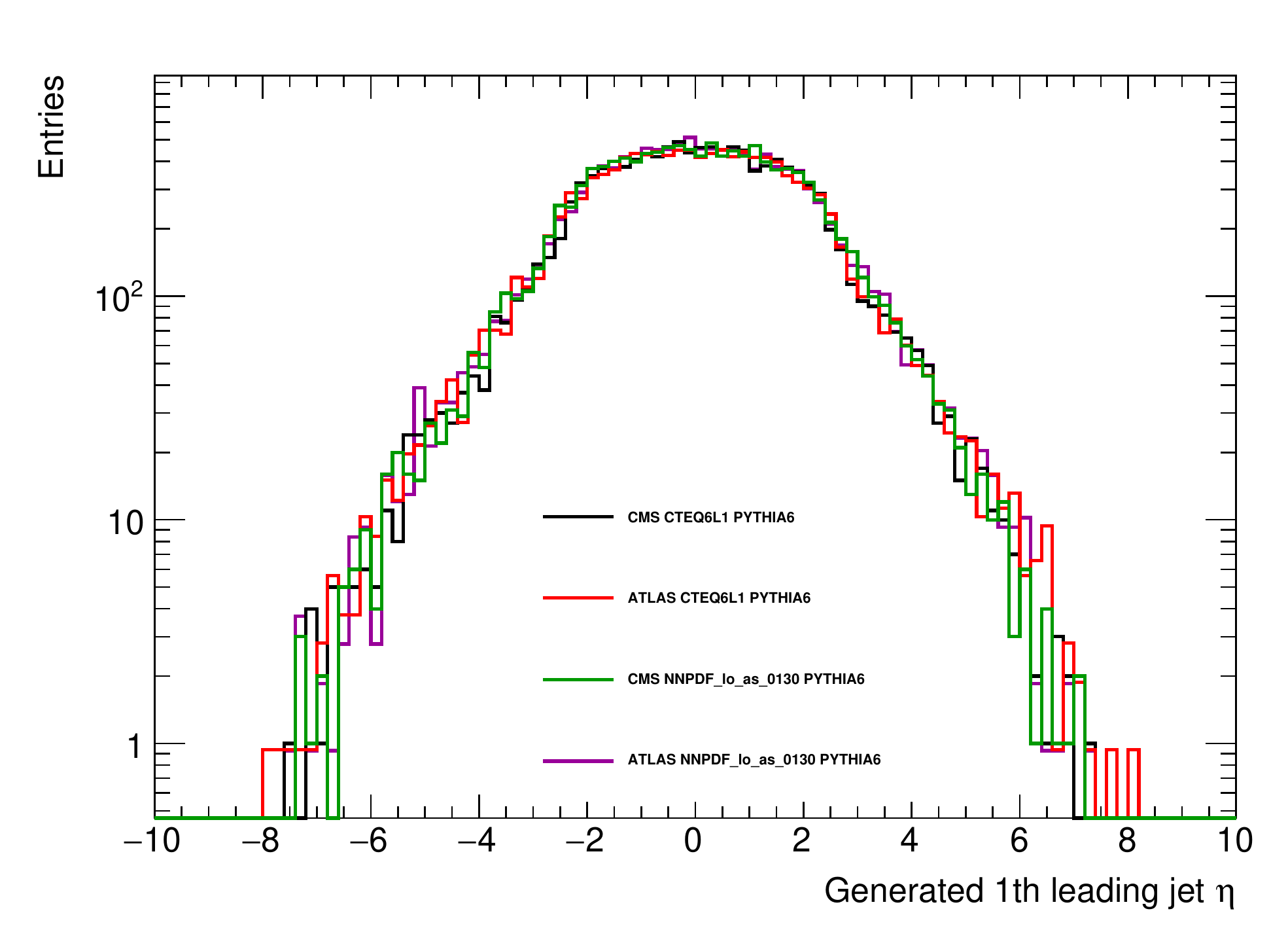}
	\caption{First leading jet $\eta$ distribution at particle level}
	\label{fig:fig6}
\end{figure}

\begin{figure}[ht]
\begin{subfigure}{0.5\textwidth}
\includegraphics[width=0.9\linewidth, height=9cm]{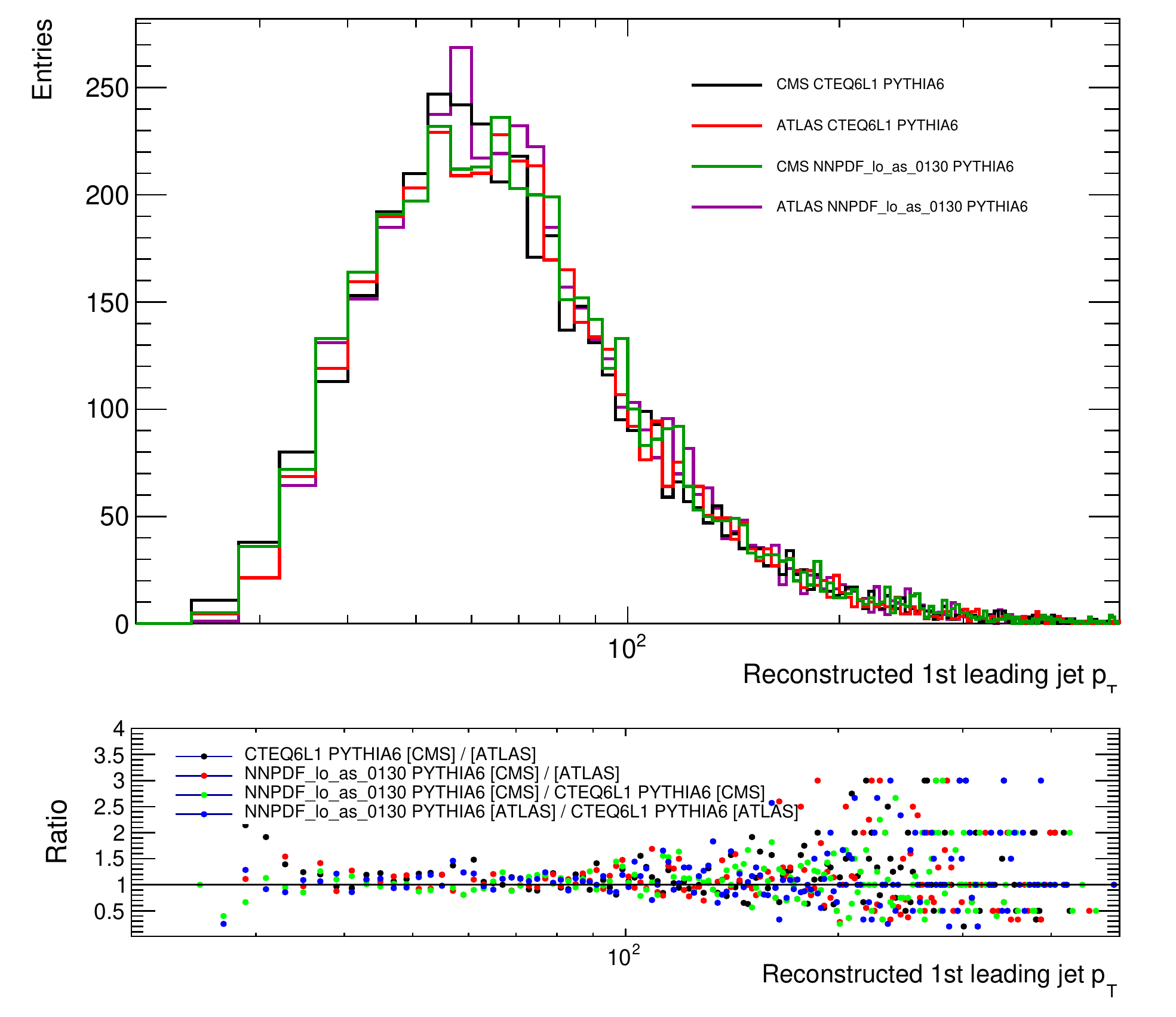}
\caption{}
\label{fig:subim7}
\end{subfigure}
\begin{subfigure}{0.5\textwidth}
\includegraphics[width=0.9\linewidth, height=9cm]{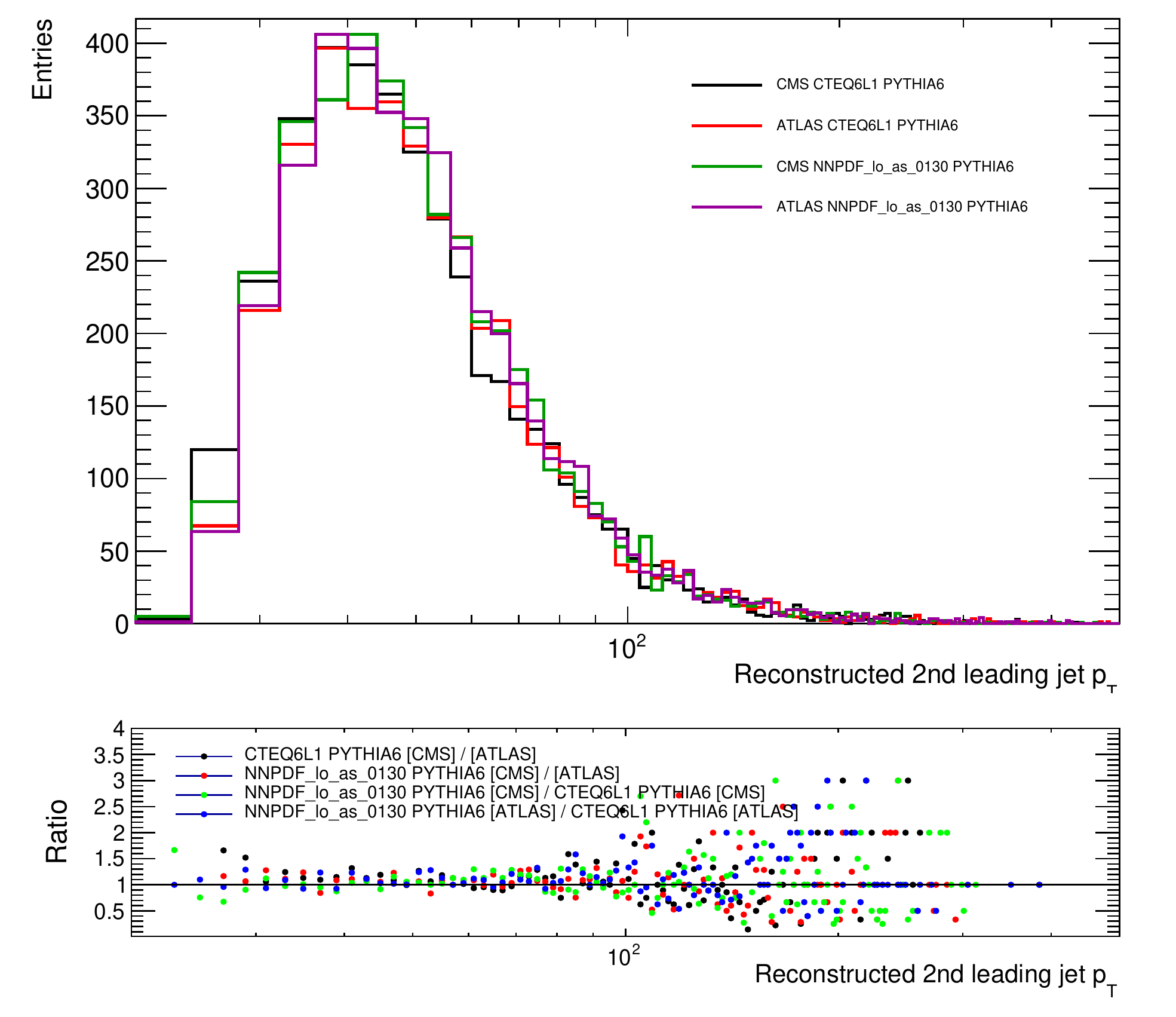}
\caption{}
\label{fig:subim8}
\end{subfigure}
\caption{(a)First leading jet $p_{T}$ for reconstructeed level which are reconstructed anti-$k_{T}$ algorithms with cone size $R= 0.8$ (b)Second leading jet $p_{T}$ for particle level}
\label{fig:fig7}
\end{figure}

\begin{figure}[ht]
	\centering
        \includegraphics[width=9cm]{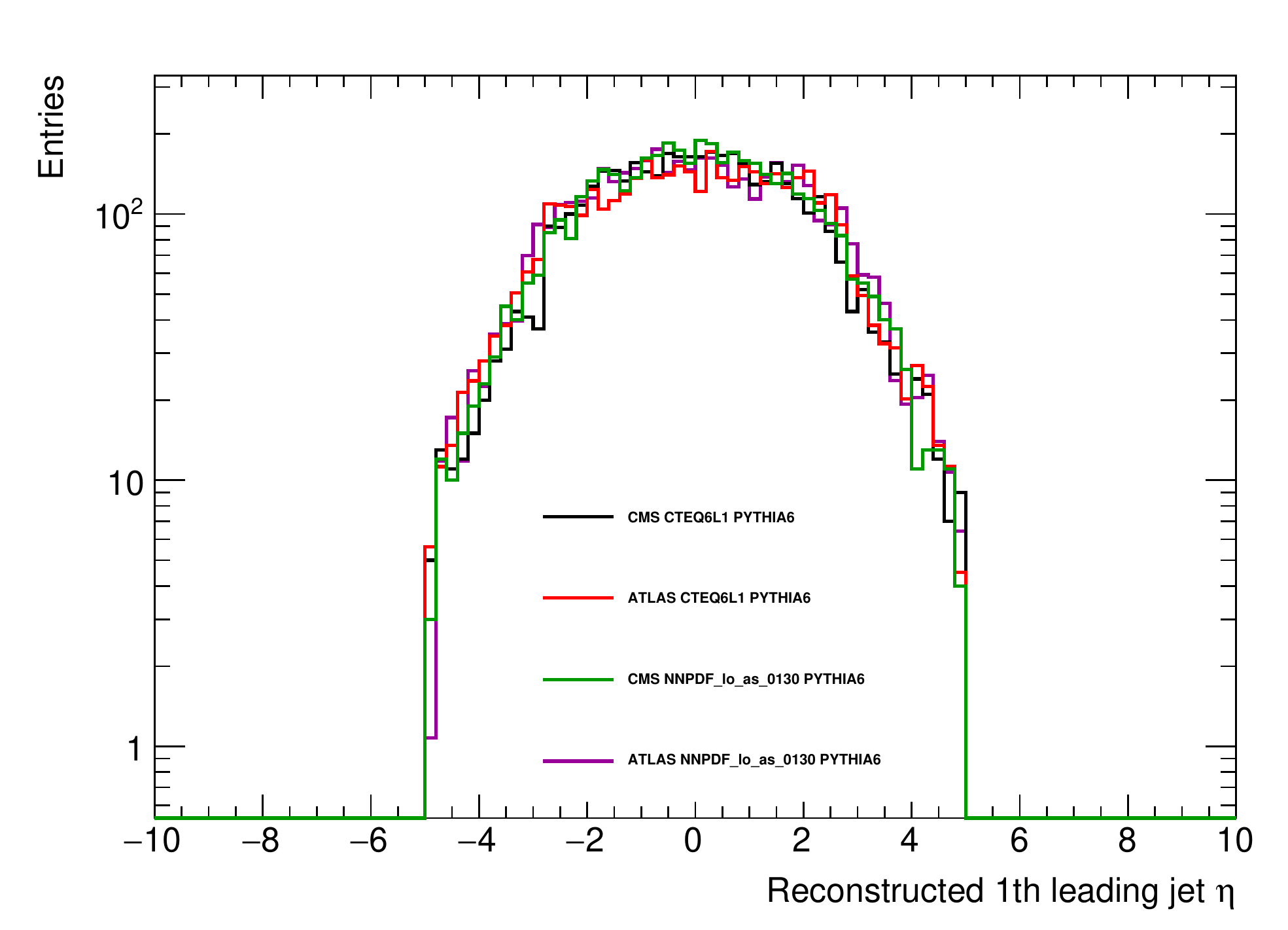}
	\caption{First leading jet $\eta$ distribution at reconstructed level}
	\label{fig:fig8}
\end{figure}

\section{Conclusion and Outlook}
$W_{L}W_{T}jj$ events are generated at $\sqrt{s}$ = 14 TeV  pp collision and corresponding to an integrated luminosity of 3000 $fb^{-1}$ at Run-III scenario for both signal and background. PYTHIA6.4 and PYTHIA8.3 are used as MC generation for parton showering, hadronization and producing nonstable hadrons decay the final particles. ATLAS and CMS official parton distribution functions (PDFs) CTEQ6L1 and NNPDF30 are used  respectively. ROOT TMVA program is used for traditional BDT multivariate discriminator, that $m(jj)$  ,$|{\Delta\eta}(jj)$, $m(W^{\pm}W^{\pm})$, zeppenfeld of two W boson, ratio between the transverse momentum of the jet system and the scalar $p_{T}$ sum of jets are chosen as inputs. After applying BDT, $W_{L}W_{T}jj$ cross-section result is improved $0.5-1.0{\%}$ for both CMS and ATLAS experiments at  $\sqrt{s}$ = 14 TeV  pp collision 3000 $fb^{-1}$ at Run-III scenario. 
\clearpage

\bibliographystyle{unsrt}  
\bibliography{references}

\end{document}